\documentclass[aps, singlecolumn, showpacs]{revtex4-2}
\usepackage{amsmath}
\begin{document}
\title {Comment on "Quantum mechanics based on real numbers: a consistent description"}

\author{S. C. Tiwari \\
Department of Physics, Institute of Science, Banaras Hindu University, Varanasi 221005, and \\ Institute of Natural Philosophy \\
Varanasi India\\ Email address: $vns\_sctiwari@yahoo.com$ \\}
\begin{abstract}
In this comment on arXiv:2503.17307 (PRL, 136, 240202, 2026), a novel idea using flag states to
transform complex state vector and operators to real ones is argued to effectively lead to a quantum
mechanics that is complex in the disguised form of real. The claim that real-valued quantum
mechanics cannot be falsified is conceptually flawed. A new criterion based on the geometric phases is suggested to compare real and complex versions of quantum mechanics.
\end{abstract}
\maketitle

Real quantum mechanics (ReQM) based on real numbers in contrast to the standard QM where complex Hilbert space and operators are fundamental, let us term it CQM, has been of interest in recent years primarily inspired by quantum information science (QIS) paradigm. In a new paper \cite{1} authors claim in the abstract that "real-valued quantum mechanics cannot be falsified, and therefore the case of complex numbers is a matter of convenience". In the present comment, such a sweeping conclusion is shown to be conceptually flawed, and the postulated construction is argued to be unfruitful. 

Conceptual flaw could be first explained by an example: real number continuum includes integers, rational numbers and irrational numbers. A set of integers or rational numbers is a sub-set of real numbers, but focusing only on either does not imply that irrationals are unnecessary or non-existent. Moreover, the set of integers has discreteness, therefore it is distinct but not false. The falsifiable ReQM, for example, \cite{2} based on real Hilbert space and real representation of the CQM postulates is only one of the possible versions of ReQM; the novel construction \cite{1} is another version but it cannot rule out further new ReQM constructs or replace CQM. A fruitful theory is one that has potential for generalization and new prediction. For example, Minkowskian spacetime geometry is equivalent to pseudo-Euclidean geometry with real space and time coordinates. However, the later one generalized to pseudo-Riemannian geometry proved fruitful to develop general relativity. On the other hand, the proposed ReQM \cite{1} merely mimics aspects of CQM. 

The papers on ReQM accept the QIS paradigm, for example, as articulated by Bub \cite{3} that "the puzzling features" of QM should be seen as resource not a problem. Bell-like experiments for Bohm's version of EPR serve the purpose of validation or falsification of the theory. However, a theory of physics that claims to describe the objective physical world must be free of mysticism/mysteries \cite{4}. This paper \cite{1} as well as its well-reasoned critique \cite{5} too follow the same path and do not throw any light on the mysteries of QM. Bell \cite{6} used 2D Hilbert space (for spinor) and noted that "space, time and causality" are absent in the hidden variable model for this system; in section VI a brief comment using $\psi_{ij}({\bf r}_1, {\bf r}_2) $ is made. Unfortunately, this crucial issue is hardly paid any attention \cite{7}. Therefore, the debate on CQM versus ReQM including \cite{1} evades the fundamental questions that bothered founders of QM.

There are two important novel technical postulates: (1) mapping of CQM to ReQM using flag states, and (2)  the replacement of the tensor product for multipartite system by the quotient with the kernel of the inverse of the map from complex to real. A critique on both \cite{5} is interesting, however I offer a differing new perspective. A complex number $z=a+ i b, ~ i=\sqrt{-1}$ can be represented as an ordered pair of real numbers $(a, b)$.  A complex state vector $|\Psi> =Re |\Psi>+ i Im |\Psi>$  in real form would be simply $(Re |\Psi>, ~ Im |\Psi>)$. Authors \cite{1} introduce flag states for a map
\begin{equation}
S: |\Psi > ~\mapsto Re( |\Psi >) \otimes | 0>_F + Im ( |\Psi >) \otimes | 1>_F =: |\tilde{\Psi}>
\end{equation}
Normalization of the state vector and observables as expectation values of the corresponding Hermitian operators in CQM imply that there exists a global phase invariance, termed as $U(1)$ ambiguity in \cite{1}. Mapping it to real
\begin{equation}
S: e^{i\alpha}|\Psi > ~\mapsto Re( |\Psi >) \otimes (\cos \alpha | 0>_F + \sin \alpha   | 1>_F)+ Im ( |\Psi >) \otimes (- \sin \alpha | 0>_F + \cos \alpha  | 1>_F)
\end{equation}
In ReQM Eq.(2) represents $SO(2)$ ambiguity; a peculiarity of this ambiguity is that at $\alpha =\pi /2$ real representation of $i  |\Psi >$ and $  |\Psi >$ are orthogonal. Transformation of the operators $\Pi$ from CQM to ReQM $\tilde{\Pi}= \mathcal{T} (\Pi)$ is 
\begin{equation}
\mathcal{T} (\Pi) = Re \Pi \otimes I_F + Im \Pi \otimes J_F
\end{equation}
where
\begin{equation}
I_F = \begin{bmatrix} 1 & 0 \\ 0 & 1 \end{bmatrix}; ~ J_F =   \begin{bmatrix} 0 &  -1 \\ 1 & 0 \end{bmatrix}
\end{equation}

The main problem is with the role of flag states and ambiguity. Eq.(1) and Eq.(8) in \cite{1} "resemble" entangled state but according to the authors these are not, and "the flag is not a directly accessible degree of freedom".  Flag states are auxiliary and if inaccessible it can be argued that these effect the single state $|\tilde{\Psi}>$ via invisible entanglement as reflected in the following: flag rotation helps implement $U(1)$ ambiguity for representation (1) and also for multipartite equivalence, and gives a formal symmetric wavefunction look to Eq.(8). Closer examination of Eq.(1) shows that two real degrees of freedom of flag $|0>_F,~|1>_F$  imprint CQM on ReQM, and the hidden significance of $i$, if any gets lost. Thus ReQM in \cite{1} is CQM in disguised form. The role of flag states and ambiguity is elaborated further.

 {\bf P1}:  Let us re-examine Eqs.(1) and (2). The complex number $z$ could be re-written as 
\begin{equation}
z \equiv a \begin{bmatrix} 1  \\ 0 \end{bmatrix} + b \begin{bmatrix} 0  \\ 1 \end{bmatrix}
\end{equation}
In classical information $ \begin{bmatrix} 1  \\ 0 \end{bmatrix}, ~  \begin{bmatrix} 0  \\ 1 \end{bmatrix}$ represent a bit, promoting these to qubit $|0>,~|1>$ implies that these are orthogonal quantum states and the superposition and entanglement give them typical QIS characteristics. If we compare this with Eq.(1) either quantum theory has to be coded in the flag states or the real numbers $(Re |\Psi>,~Im |\Psi>)$ in Eq.(1) make it a trivial mapping of complex to real ordered pairs. In fact, geometrical representation of complex number $z$ in the Argand diagram shows that so-called $U(1)$ ambiguity corresponds to the nonsimply connected circle $S^1$. The mapping $i \mapsto J_F$ transforms $e^{i\alpha} \rightarrow R(\alpha)$, and the group space of $SO(2)$ is also a circle. Note that in 2D the basis vectors $ \begin{bmatrix} 1  \\ 0 \end{bmatrix}, ~  \begin{bmatrix} 0  \\ 1 \end{bmatrix}$ are orthogonal, and the rotation matrix $R(\alpha)$ at $\alpha =\pi/2$ is just $J_F$ that rotates $ \begin{bmatrix} 1  \\ 0 \end{bmatrix}$ to orthogonal  $\begin{bmatrix} 0  \\ 1 \end{bmatrix}$.  In the context of QM \cite{1}, what is the import of the statement "the flag is not a directly accessible degree of freedom"? Bang et al \cite{5} make an important point that in QIS flag has physical significance, however in \cite{1} it is analogous to a gauge coordinate thus CQM is represented by a real gauge-redundant theory. Another viewpoint could be that Eq.(1) represents "invisible entanglement" in analogy to black hole quantum physics: in a  pure bi-partite
system of inside and outside of a black hole, measurement on the subsystem outside results into a mixed state that has imprint of the information of the invisible black hole. It seems the idea of flag states is misleading.

{\bf P2}: Let us consider a real Hamiltonian with real eigenfunction \cite{9}
\begin{equation}
H_w =  \begin{bmatrix} x & y \\ y & -x \end{bmatrix} ; ~\Psi_w=\begin{bmatrix} \cos \phi/2  \\ \sin \phi/2 \end{bmatrix}
\end{equation}
Energy eigenvalues are $\pm r$. As $\phi$ is varied from $0$ to $2\pi$ the sign of $\Psi_w$ reverses, whereas a valid complex solution $e^{i\phi/2}~\Psi_w$ is invariant under $\phi \rightarrow \phi +2\pi$. Application of the construction prescribed in \cite{1} to the complex eigenfunction leads to the real eigenfunction $[\cos^2\phi/2,~ \sin \phi/2 \cos \phi/2,~  \sin \phi/2 \cos \phi/2, \sin^2 \phi/2]^T$ that contradicts the solution $\Psi_w$ that becomes $[\cos \phi/2,~0,~\sin \phi/2,~0]^T$. A single example is sufficient to prove that the main claim made in \cite{1} is invalid.

{\bf Nontrivial topology:}  The sign ambiguity in the above example is resolved \cite{9} calculating the geometric phase that is determined by an Aharonov-Bohm like gauge potential $(-i/2)$. A new insight is gained that the analogue of geometric phases in CQM for ReQM could be a significant criterion to compare the two. Recently, I have obtained a radically new result in ReQM that $i$ hides spin \cite{10}. The important significance of this example discussed in \cite{9, 10} is that the issue of geometric phases in CQM and ReQM could be a new test to prove the equivalence or inequivalence of the real and complex versions of quantum mechanics. It could throw light on the claim that experimentally ReQM and CQM are indistinguishable \cite{11}; one may also examine a broader picture of real wave equations \cite{12}.

 {\bf Acknowledgment}: I acknowledge correspondence with J. Bang, P. B. Hita and O. Passon.

\end{document}